\documentclass[showpacs,twocolumn,floatfix,amssymb,amsmath,aps,prb]{revtex4}
\newif\ifpdf
\ifx\pdfoutput\undefined
   \pdffalse              
\else
   \pdfoutput=1
   \pdftrue
\fi  

\ifpdf
  \usepackage[pdftex]{graphicx}
\else
  \usepackage[dvips]{graphicx}
\fi

\ifpdf
   \RequirePackage[pdftex, pdfpagemode=none, pdftoolbar=true,
                pdffitwindow=true, pdfcenterwindow=true]{hyperref}
   \usepackage{thumbpdf} 
\else
\fi

\usepackage{bm,times} 
\begin{document}


\newcommand{\phdagg}{^{\phantom{\dagger}}}
\newcommand{\phid}{\phi^{\dagger}}
\newcommand{\phipd}{\phi^{\phantom{\dagger}}}
\newcommand{\vpi}{{\boldsymbol{\pi}_{\bf{a}}}}
\newcommand{\order}[1]{\mathcal{O}(#1)}
\newcommand{\mb}[1]{\mathbf{#1}}
\newcommand{\cscucl}{Cs$_2$CuCl$_4$\ }
\newcommand{\be}{\begin{equation}}
\newcommand{\ee}{\end{equation}}
\newcommand{\s}{{\bf{S}}}
\renewcommand{\r}{{\bf{R}}}
\renewcommand{\d}{{\boldsymbol{\delta}}}
\newcommand{\OO}{{\bf{0}}}
\newcommand{\Q}{{\bf{Q}}}
\newcommand{\QC}{{\bf{L}}}
\newcommand{\q}{{\bf{q}}}
\newcommand{\qp}{{\bf{q^{\prime}}}}
\renewcommand{\k}{{\bf{k}}}
\newcommand{\kp}{{\bf{k^{\prime}}}}
\newcommand{\ku}{{{\bf{M}}_{1}}}
\newcommand{\kd}{{{\bf{M}}_{2}}}
\newcommand{\vd}{{\bf{\Delta M}}}
\newcommand{\p}{{\bf{p}}}
\newcommand{\pp}{{\bf{p^{\prime}}}}
\renewcommand{\b}{{\bf{B}}}
\newcommand{\h}{{\bf{H}}}
\newcommand{\hh}{H}
\newcommand{\D}{{\bf{D}}}
\newcommand{\ham}{{\mathcal{H}}}
\newcommand{\epsl}{\epsilon}

\title{Commensurate and incommensurate ground states of 
Cs$_2$CuCl$_4$ in a magnetic field} 
\author{ M.\ Y.\ Veillette and J.\ T.\ Chalker}
\affiliation{Theoretical Physics, University
of Oxford, 1, Keble Road, Oxford, OX1 3NP, United Kingdom\\}
\date{\today} 

\begin{abstract}

We present calculations of the magnetic ground state of \cscucl 
in an applied magnetic field, with the aim of understanding
the commensurately ordered state that has been 
discovered in recent experiments. 
This layered material is a realization of a Heisenberg antiferromagnet 
on an anisotropic triangular lattice. Its behavior in a magnetic field
depends on field orientation, because of weak Dzyaloshinskii-Moriya interactions.    
We study the system by mapping the spin-$1/2$ Heisenberg Hamiltonian onto a
Bose gas with hard core repulsion. This Bose gas is dilute, and calculations are controlled, 
close to the saturation field. We find a zero-temperature  transition between 
incommensurate and commensurate
phases as longitudinal field strength is varied, but only incommensurate
order in a transverse field. Results for both field orientations are consistent with
experiment.

\end{abstract}
\maketitle
\section{Introduction}

Experiments on the spin $S=1/2$ triangular lattice antiferromagnet \cscucl have
mapped out its properties in great detail over the last few 
years.~\cite{Coldea00, Coldea1, Coldea2, Coldea3, Coldea4, Unknown01} It has attracted particular
interest because its low-dimensionality, frustrated interactions and small spin
are all features expected to promote quantum fluctuations.
It is therefore a system for which long-standing proposals~\cite{Anderson01} of
quantum disordered states may be relevant at intermediate
energy scales, despite the fact that it has conventional N\'eel order
at the lowest temperatures.
Experimental studies are facilitated by the small strength of exchange interactions,
which makes it possible to reach the saturation magnetization with accessible fields
and also results in relatively low excitation energies. Unusual
features of the excitations revealed in inelastic neutron scattering~\cite{Coldea3}
have attracted
considerable theoretical
interest.\cite{Bocquet01,Chung01,Chung02,Isakov01,Veillette02,Alicea01,Zheng03}
 
The magnetic phase diagram of \cscucl as a function of temperature,
magnetic field strength and field orientation is remarkably rich:
it reflects the competition between exchange, Dzyaloshinskii-Moriya (DM)
and Zeeman energies. In zero field, incommensurate spiral long range order is observed
at temperatures below $T_N=0.62 {\rm K}$. The ordered
moments lie in an easy plane (the $b$-$c$ plane, see Fig. \ref{Figure1}) 
selected by the DM interactions. An applied magnetic field 
has different effects depending on whether it is transverse
to this plane (along the $a$-axis) or longitudinal (within the $b$-$c$ plane).
In a transverse field at low temperature,
ordered moments cant out of the easy plane, lying on a cone,
and an incommensurately ordered phase is observed for all field 
strengths up to the saturation field of $B^a_{cr}=8.51{\rm T}$.
The magnetization, incommensurate ordering wavevector and
transverse order parameter in this phase have, we believe,
been well accounted for in earlier theoretical work by the present 
authors in collaboration with Coldea,
using the $1/S$ expansion, together with dilute 
Bose gas techniques close to the saturation field.~\cite{Veillette01} 
Thermal effects in the dilute Bose gas have been
examined recently in Refs.~\onlinecite{Coldea4} and \onlinecite{Kovrizhin}.

By contrast, behavior in longitudinal fields is less well
understood. From a theoretical viewpoint, competition between
magnetic field and easy plane anisotropy adds
complexity to the problem. Using a classical (large $S$)
calculation, we found in earlier work~\cite{Veillette01}
two incommensurate low-temperature phases: 
a distorted cycloid spin structure in low fields, and a cone state
at higher fields, with a ferromagnetically aligned state
above the saturation field.
Initial neutron diffraction studies in a magnetic field 
directed along the
$c$-axis identified a distorted cycloid 
for fields below $B^c<1.4 {\rm T}$
and elliptical incommensurate order 
for $1.4 {\rm T}< B^c<2.1 {\rm T}$.\cite{Comment1}
Subsequent measurements
detected  a further incommensurate
phase in a narrow range just below 
the saturation field $B^c_{cr}=8.02{\rm T}$,
for $7.1{\rm T}<B<B^c_{cr}$ (Ref.\onlinecite{Veillette01}),
and most recently,
intervening commensurate order, 
for $2.1{\rm T}<B^c<7.1{\rm T}$ (Ref.\onlinecite{Unknown01}).
All incommensurate phases have magnetic Bragg peaks at the
field-dependent wavevectors $\bm{Q}(B)=(0,Q(B),0)$, whereas in the commensurate
phases Bragg peaks are observed on the Brillouin zone boundary, at
$\ku, \kd =(0,\frac{1}{2},\pm \frac{1}{2})$.

In the present paper, we examine theoretically the zero temperature 
phase diagram of \cscucl in a magnetic field with a strength close
to the saturation value. We focus on the observed differences in the effects
of longitudinal and transverse fields.
Behavior close to the saturation field is of special interest because 
in this regime quantum fluctuations are under theoretical control,
even for $S=1/2$. Reversed spins form a gas of hard core bosons
which is dilute, and interactions between bosons can be treated exactly at low
density by summing two-body scattering processes to obtain an effective
interparticle potential.~\cite{Batyev01, Batyev02, Gluzman01, Nikuni01} 
Using this approach we have shown 
previously~\cite{Veillette01} that the system has an incommensurate
cone ground state at fields just below the saturation value, for both
field orientations. In the following we add to this with the
finding that the dilute Bose gas treatment
yields a first order quantum phase transition to a commensurately ordered
ground state below a longitudinal field strength of $B_{T1}^c=0.96 B^c_{cr}$.
Since commensurate order appears at a field strength $B_{T1}$ which is close
to the saturation field $B^c_{cr}$, it takes place under conditions
for which our calculations are expected to be reliable.
For a system in a transverse field the same method gives only an incommensurate
cone state. Results for both field orientations are therefore consistent
with experiment. Our ideas can be further tested experimentally 
by comparing the magnetic structure
we predict for the commensurate phase with observations.

\section{Model and Calculations}

\begin{figure}[ht]
\begin{center}
\includegraphics[width=6.5cm]{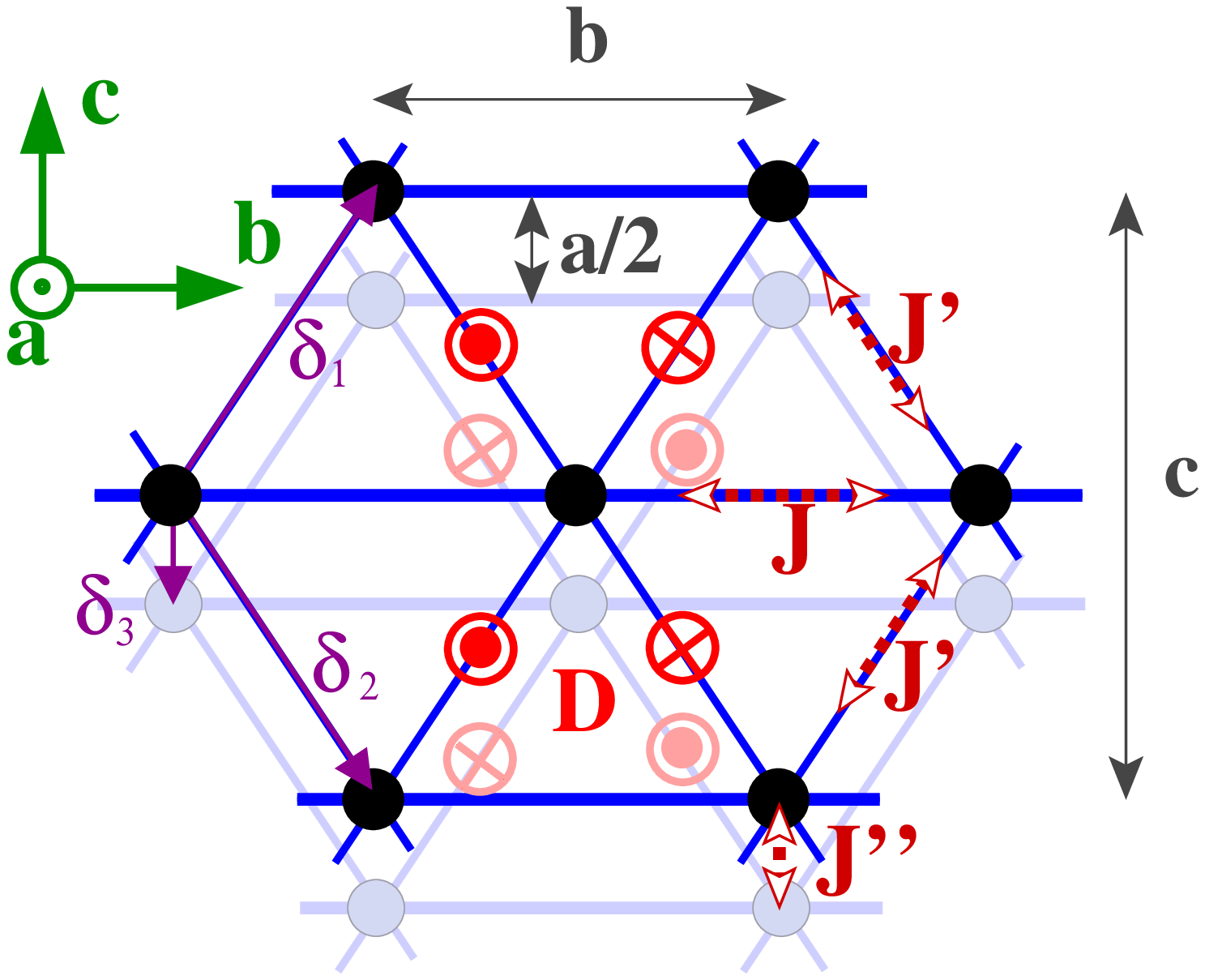}
\includegraphics[width=4.25cm]{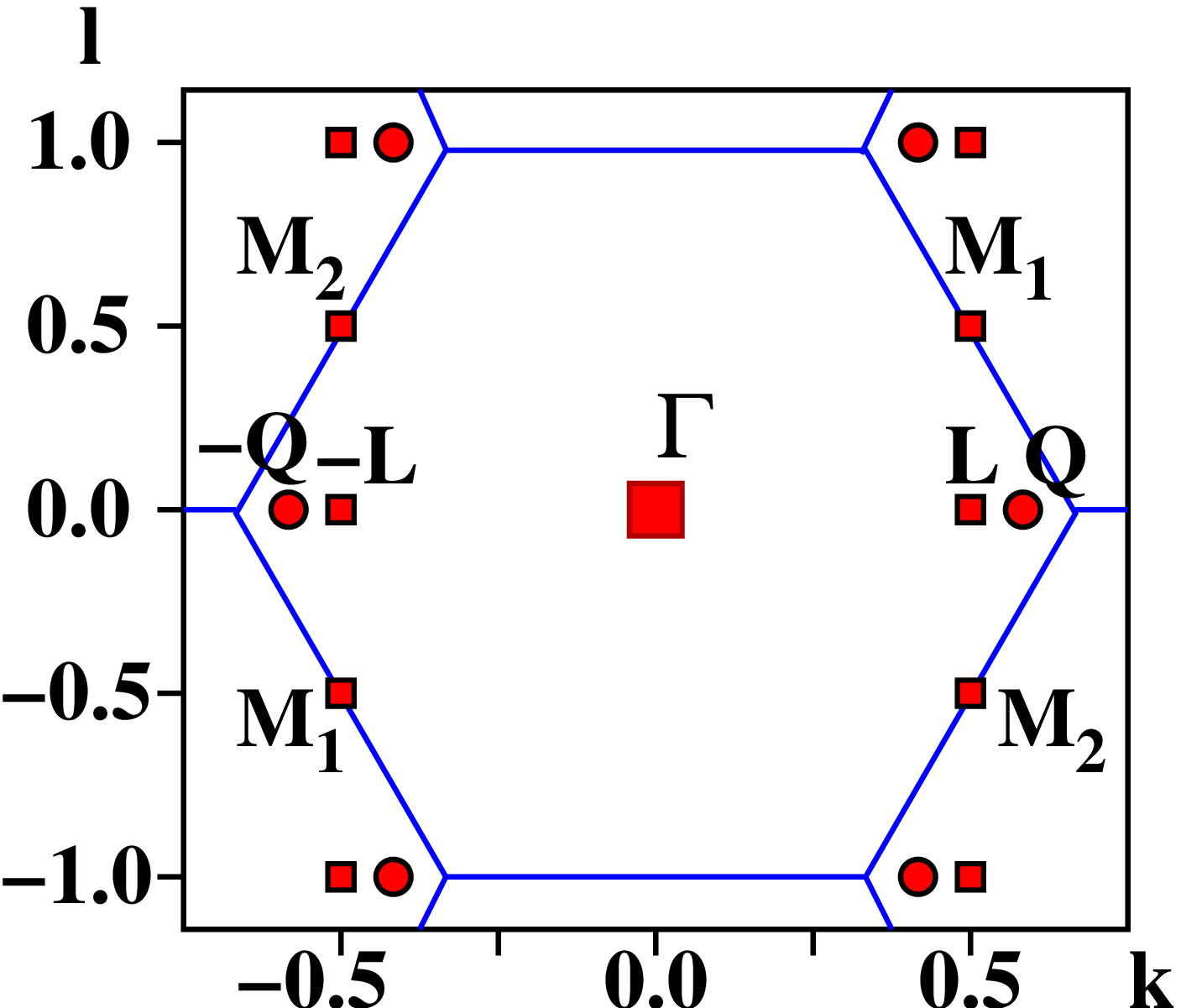}
\includegraphics[width=4.25cm]{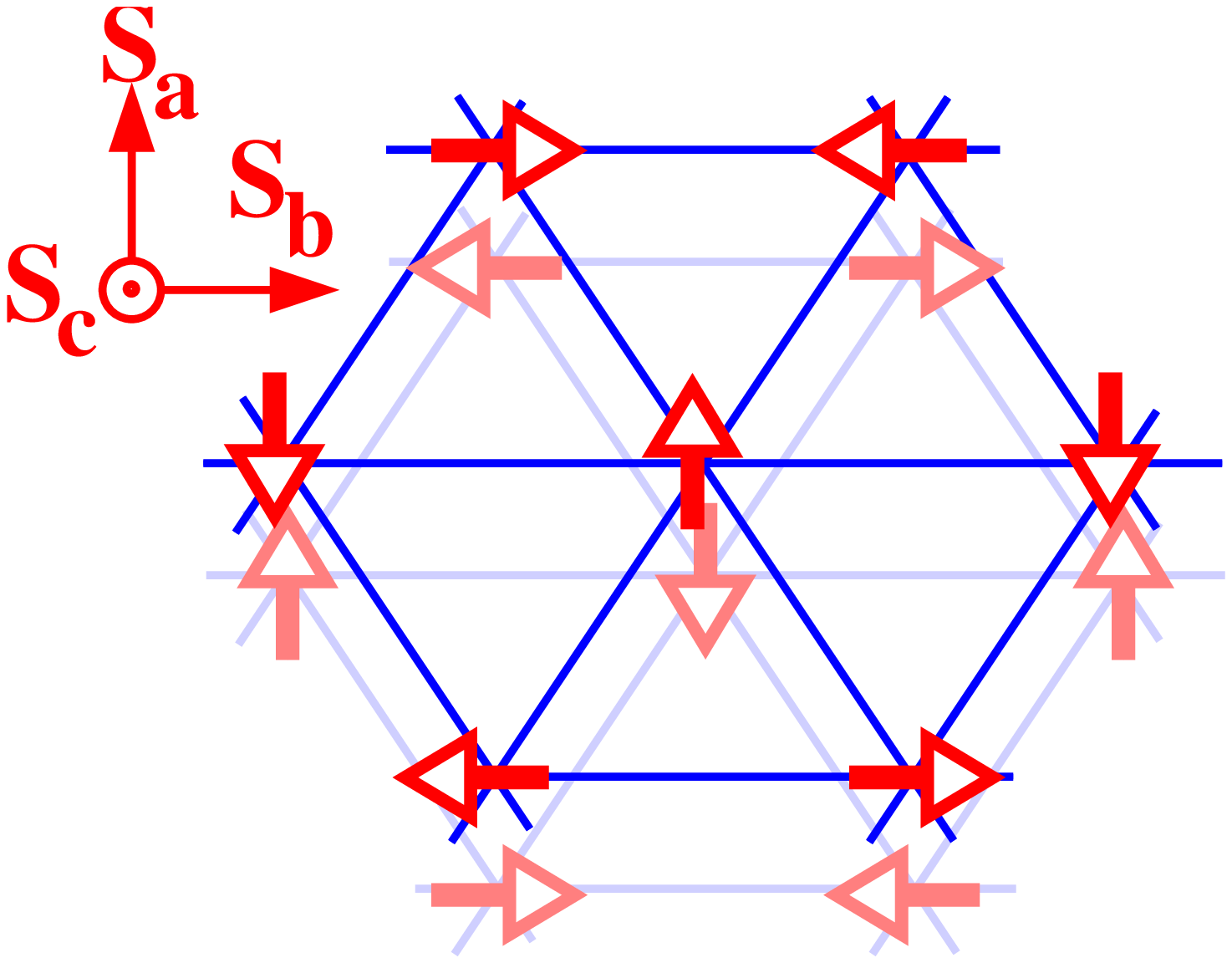}
\caption{(Color Online) (A) Magnetic sites and exchange couplings of
\cscucl: Even (color) and odd (gray) layers are stacked along the
$a$-direction. 
The signs
$\boldsymbol{\otimes}, \boldsymbol{\odot}$ refer to the direction of
the DM vector between zig-zag bonds. (B) Reciprocal space plane $(k, l)$, with
the projected Brillouin zone, projections of the Brillouin
zone center $\Gamma$, the incommensurate ($\Q, -\Q$) and the commensurate
$(\QC, \ku, \kd)$ ordering wavevectors indicated. (C) 
Predicted antiferromagnetic state (Eq. \ref{SC2}) with Bragg
peaks at $\ku$ and $\kd$, showing the orientation  of ordered moments projected onto
the $a$-$b$ spin plane. Note that here, of necessity, spin coordinates
are rotated relative to lattice coordinates.}
\label{Figure1}
\label{Figure3}
\end{center}
\end{figure}

To proceed, we first recall the Hamiltonian appropriate for
\cscucl, which has been determined with precision by
inelastic neutron scattering measurements of 
the energies of single spin-flip excitations
from the fully polarized, high field state.~\cite{Coldea2} 
Magnetic moments in a single layer of the material lie
on sites of an anisotropic triangular lattice 
as shown in Fig.~\ref{Figure1}.
The dominant exchange interaction $J=4.34{\rm K}$
is along the crystallographic $b$ axis. A weaker
exchange $J^{\prime}=0.34 J$
acts on the zig-zag bonds, and a much weaker interaction
$J^{\prime \prime}=0.045 J$ couples layers. DM interactions
are symmetry-allowed on the
zig-zag bonds. The spin Hamiltonian can be written as
$\ham=\ham_0+\ham_{DM} + \ham_{B}$,
where $\ham_0$, $\ham_{DM}$ and $\ham_{B}$ include the isotropic, DM
and Zeeman interactions, respectively. Here
\be 
\ham_0 = \sum_{\r, \d} J_{\d} \s_\r \cdot \s_{\r+\d}, 
\label{ham0}
\ee
where $\d$ denotes bond vectors connecting neighboring sites 
and $J_{\d}$ represents $J$, $J^{\prime}$ or $J^{\prime \prime}$, 
as shown in Fig.~\ref{Figure1}. The
DM interaction is
\be 
\ham_{DM} = \sum_\r \! \D^{\pm} \! \cdot \s_\r \times \left[
\s_{\r+\d_1} \! + \! \s_{\r+\d_2} \!
\right].
\label{hamDM}
\ee 
The vector $\D^\pm=(\pm D, 0, 0)$ (with $D=0.053 J$) is
perpendicular to the layers and alternates in sign between
even and odd layers because they are
inverted versions of one another. The Zeeman interaction is
$\ham_{B}= - \sum_\r \h \cdot \s_\r,$ 
where $\h$ is a reduced Zeeman field
with components $H^{\alpha} = g^{\alpha}\mu_B B^{\alpha}$.

This spin model can be treated as a lattice gas of hard core 
bosons.~\cite{Matsubara01, Batyev01, Batyev02, Gluzman01, Nikuni01}
In the dilute limit, all quantum effects are incorporated exactly by
summing ladder diagrams for the
interaction vertices.~\cite{Beliaev01, Abrikosov} A Hartree Fock treatment
of these effective interactions can be used to obtain the ground state spin structure.  We describe 
further details of the calculation separately for the two cases of longitudinal and
transverse magnetic fields.
\subsection{Longitudinal Fields}

For definiteness, we consider a field
in the $c$ direction. We introduce
boson creation and annihilation operators $\phid_\r $ and $\phipd_\r
$ and set $S^{c}_\r = \frac{1}{2} - \phid_\r
\phipd_\r $, $S^{+}_\r = S^{a}_\r +i S^{b}_\r = \phipd_\r $ and
$S^{-}_\r = S^{a}_\r -i S^{b}_\r = \phid_\r$.
For the spin commutation relations to be satisfied, a hard core
constraint on boson number must be imposed by including an
on-site repulsion $U$ in the Hamiltonian and taking the limit $U\to \infty$.

The DM interaction is represented by terms cubic in
boson creation and annihilation operators, because
for this field orientation it couples transverse and longitudinal spin
components. Close to the saturation field we are concerned 
ultimately with an effective description involving only bosons 
near minima of the dispersion
relation generated by the exchange coupling.
As argued previously, \cite{Veillette01} momentum
conservation precludes cubic terms in an effective low energy Hamiltonian.
Instead, within second order perturbation theory
the DM interactions generate quadratic and quartic 
couplings between low energy bosons, with magnitude ${\cal O}(D^2/J)$.
Since $\left| D/J \right| \ll 1$, we neglect altogether the effect of DM
interactions in {\em longitudinal} fields.

At this point a simplification is possible. 
In the absence of DM interactions the distinction between odd and even
layers disappears. The size of the unit cell in the $a$-direction is halved
and that of the Brillouin zone is doubled. We take Fourier transforms using
this reduced unit cell.
The Hamiltonian (omitting a constant) is
\begin{align}
\ham &= \sum_\k (\epsl_\k -\mu)\phid_\k \phipd_\k + \frac{1}{2 N}
\sum_{\k, \kp, \q} V_\q \phid_{\k+ \q} \phid_{\kp- \q} \phipd_\kp
\phipd_\k,
\end{align} 
where
$\epsl_\k = J_\k -J_{\rm min} $ is the boson kinetic energy,
$J_\k=1/2\sum_{\d} J_\d e^{i \k \cdot \d}$ is the Fourier transform of
the exchange couplings, $\mu =\hh^c_{cr} - \hh^c$ is the boson
chemical potential, and $\hh^c_{cr} = J_{\OO}-J_{\rm min}$ is the
saturation field. The number of lattice sites is $N$ and
the interaction vertex is given by $V_\q = 2J_\q+ 2 U $.
Writing $\k=(h,k,l)$ as shorthand for $\k=2 \pi (h/a,k/b,l/c)$,
the dispersion relation $\epsl_{\k}$ has two minima located at the
incommensurate wavevectors $\k=\pm\Q=\pm (1,1/2+\epsilon_0,0)$,
where $\sin \pi\epsilon_0=J^{\prime}/(2J)$. 
On reverting to the standard description with two layers in a unit cell, these
wavevectors are denoted by $\Q=\pm (0,1/2+\epsilon_0,0)$
and amplitudes are staggered on the two layers.\cite{Coldea2}
Note that the degeneracy of the two minima follows from symmetry
of the full Hamiltonian~\cite{Veillette01}
and is not lifted by DM interactions.


At this point we can apply standard techniques developed for the
interacting Bose gas.~\cite{Abrikosov} With negative $\mu$, the ground state
is the boson vacuum. Equivalently, with $\hh^c > \hh^c_{cr}$ the
spin system is fully polarized. With positive $\mu$, the boson
density is non-zero.
Condensation of these bosons at low temperature is
equivalent to magnetic order, and we introduce the
complex order parameter $\psi_\q=\langle \phi_\q \rangle/\sqrt{N}$. 
In the dilute regime, the full
scattering vertex can be obtained by summing pair
interactions in the particle-particle channel.
We denote this vertex, for the scattering amplitude 
of incoming bosons with momenta $\k$ and $\kp$ to outgoing states
with momenta $\k-\q$ and $\kp+\q$, by $\Gamma_{\q, \k, \kp}$.
At low density it satisfies the Bethe-Salpeter equation 
\be \Gamma_{\q, \k, \kp}\!\!=\!\! V_\q - \frac{1}{N} \sum_{\qp} \frac{
V_{\q-\qp}}{\epsl_{\k+\qp}+\epsl_{\kp-\qp}-\epsl_{\k}-\epsl_\kp }
\Gamma_{\qp, \k, \kp}\,.
\label{Gamma}
\ee
This integral equation can be reduced
to a set of linear equations which are readily solved
numerically
(see Ref.~\onlinecite{Veillette01}).

We now proceed to an analysis of the energy of various candidate ground
states. This energy can be written in the form of a Landau
expansion, in terms of one or more order parameters $\psi_\q$.
We first review our earlier results, for behavior immediately
below the saturation field.~\cite{Veillette01}
In this regime one expects condensation only at one or both
minima of the boson dispersion relation, with wavevectors $\k=\pm \Q$. 
Allowing for two possible order parameters, the energy per site is
\begin{align}
E/N &= \left(\epsl_\Q -\mu \right) \left(\left|\psi_\Q \right|^2
 +\left|\psi_{-\Q} \right|^2 \right) \notag \\ & + \frac{1}{2}
 \Gamma_{1} \left( \left|\psi_\Q \right|^4 +\left|\psi_{-\Q} \right|^4
 \right) +\Gamma_2 \left|\psi_\Q \right|^2\left|\psi_{-\Q} \right|^2 ,
\label{Energy1}
\end{align}
with $\Gamma_1=\Gamma_{\OO, \Q, \Q}=\Gamma_{\OO, -\Q, -\Q}$  and
$\Gamma_2=\Gamma_{\OO, -\Q, \Q}+\Gamma_{2\Q, -\Q,
\Q}$. (Here $\epsilon_\Q=0$, but we retain it for future reference). 
Minimizing the energy
with respect to the order parameters, a single component
state is favored for $\Gamma_1 < \Gamma_2$. It is the cone
state with spin structure $\langle \s_\r \rangle= ( \sqrt{n}_\Q
\cos(\pm \Q \cdot \r+\alpha), \sqrt{n}_\Q \sin(\pm \Q \cdot
\r+\alpha), \frac{1}{2}-n_\Q) $, where $n_\Q= (\mu-\epsl_\Q)/\Gamma_1$ is
the condensate density and $\alpha$ is an arbitrary phase. 
A two-component state with equal amplitudes
for both order parameters is favored for $\Gamma_1 > \Gamma_2$. 
In this state, ordered moments for a spin fan, with $\langle \s_\r \rangle = ( 2
\sqrt{n}_\Q \cos(\Q \cdot \r+\alpha)\cos(\beta), 2 \sqrt{n}_\Q \cos(\Q
\cdot \r+\alpha)\sin(\beta), 1/2-4 n_\Q \cos^2(\Q \cdot \r +\alpha) )
$, where $n_\Q=n_{-\Q}=(\mu-\epsl_\Q)/(\Gamma_1+\Gamma_2)$ and $\alpha,
\beta$ are arbitrary phases.
A numerical evaluation of Eq.~\ref{Gamma} yields $\Gamma_1=3.79 J$ and
$\Gamma_2= 4.67 J$, 
so the cone state is selected. It has energy per site 
${E}/{N}=-{(\epsilon_\Q - \mu)^2}/{2 \Gamma_1}$.

\begin{figure}[ht]
\begin{center}
\includegraphics[width=8.6cm]{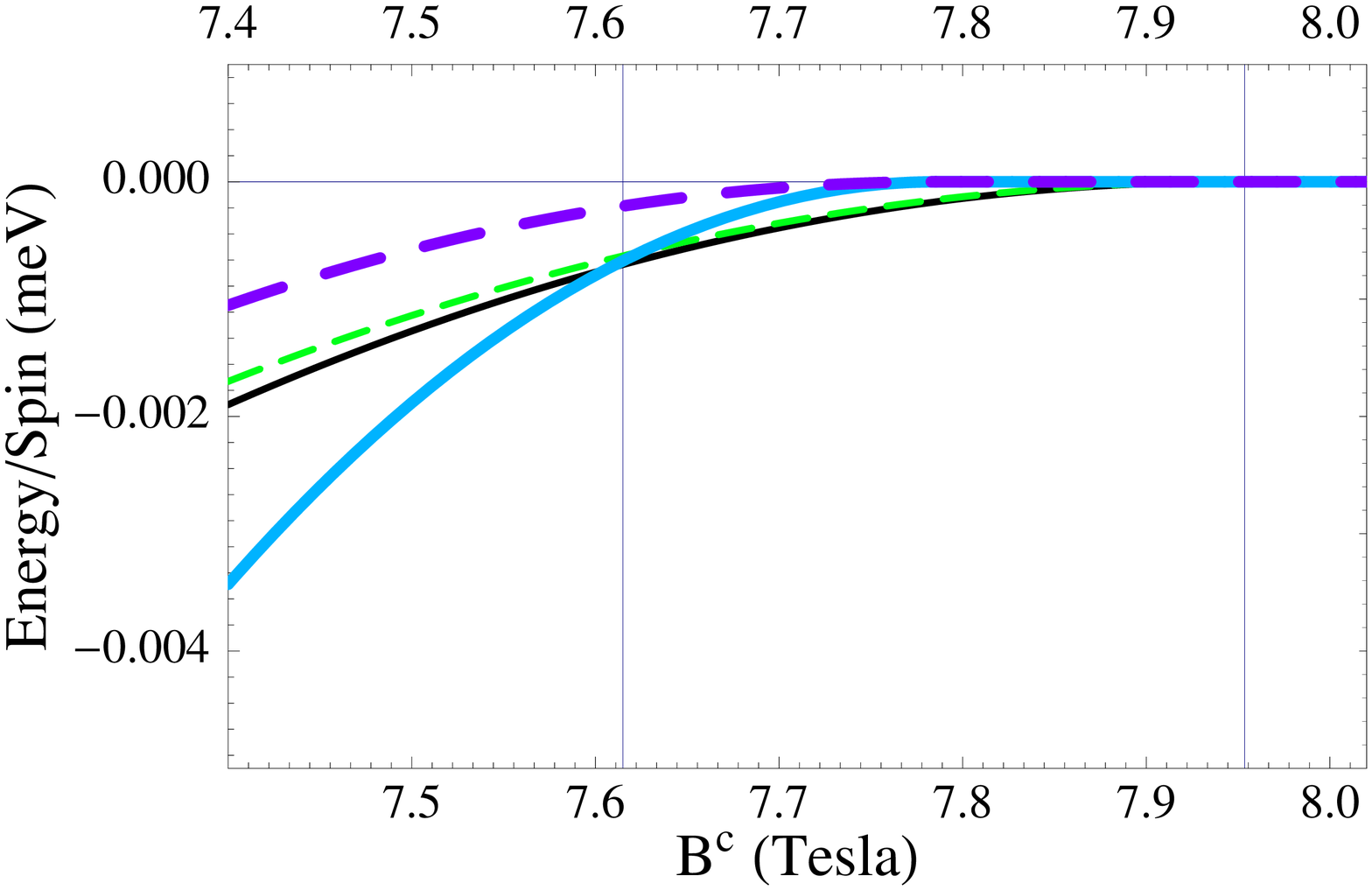}
\includegraphics[width=8.6cm]{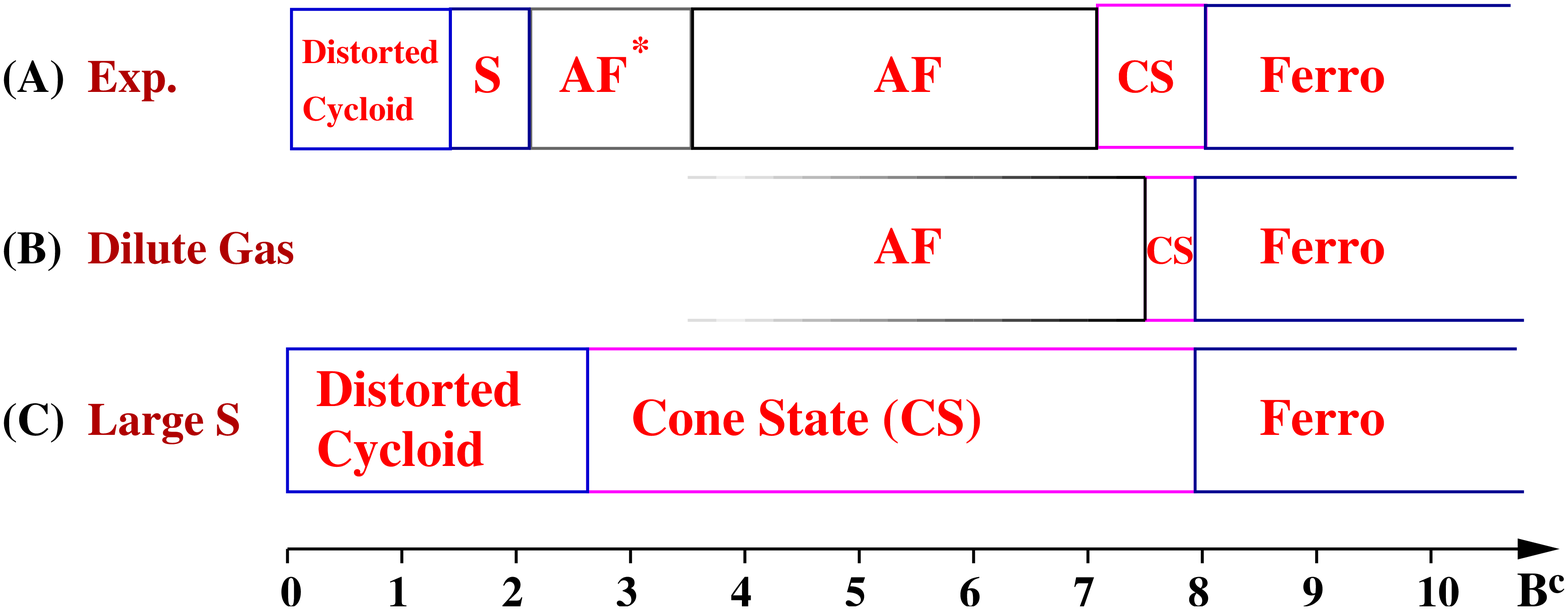}
\caption{(Color Online) Ground states in longitudinal fields. Top: 
Energies of states considered [Eqns.~(\ref{Energy1}) and
(\ref{Energy2})] as a function of longitudinal magnetic field
$B^c$, computed using $g^c=2.30$.\cite{Gfactor} 
The thin, thick, short-dashed and long-dashed lines represent,
respectively: the incommensurate cone state; the commensurate
state with condensates at $\ku$ and $\kd$; the incommensurate spin-fan; and the commensurate
state with condensate at $\k=\QC$. The incommensurate cone state is 
the ground state for $B^c_{T1}< B^c < B^c_{cr}$, with
$B^c_{cr}=7.95 T$ and $B^c_{T1}=7.615T$. The commensurate
state at $\ku$ and $\kd$ is the ground state for $ B^c < B^c_{T1}$.
Bottom: Phase diagram in longitudinal fields: 
(A) from experiment; (B) from dilute Bose gas calculation; 
(C) from large S-expansion\cite{Veillette01}. Here $S$ denotes
an incommensurate phase, and $AF$ and $AF^*$ indicate commensurate phases,
observed experimentally to have magnetic Bragg peaks at $\ku$ and $\kd$.}
%
\label{Figure4}
\end{center}
\end{figure}

At this stage, we have established that the ground state for $\mu=0^+$
is the cone state. Next we examine whether there is a transition to a commensurate 
ground state at a larger value of $\mu$. A first possible commensurate
state is one with order at the wavevector $\QC=(1, 1/2, 0)$, using our
description with a single layer per unit cell (equivalent to $(0, 1/2, 0)$
in the standard notation with two layers per unit cell).
This is a natural choice because $\QC$ lies close to $\Q$, and because
the wavevector of incommensurate order is known to move towards
$\QC$ with increasing $\mu$.\cite{Veillette01}
We find nevertheless, repeating the calculations
we have outlined, but with $\QC$ in place of $\Q$
and with $\epsilon_{\QC}> 0$, that the commensurate
spin cone and spin fan states are higher in energy than the
incommensurate states, as shown in Fig.~\ref{Figure4}.


An second possible commensurate state is one with
condensates at the wavevectors $\ku=\left(1, \frac{1}{2},
\frac{1}{2} \right)$ and  $\kd=\left(1, -\frac{1}{2}, \frac{1}{2}
\right)$, in our notation 
(or $\left(0, \frac{1}{2},\frac{1}{2} \right)$ 
and $\left(0, -\frac{1}{2}, \frac{1}{2}\right)$ 
in the standard notation), and it is in fact
at those positions that magnetic Bragg peaks are
observed experimentally.\cite{Unknown01}
The difference $\vd=\kd-\ku$ is 
half a reciprocal lattice vector so that umklapp scattering
between the condensates is allowed and may reduce the energy of the state.
We obtain an energy per site 
\begin{align}
E/N &=  \left(\epsl_{\ku} -\mu \right) \left( \left|\psi_\ku \right|^2
 + \left| \psi_\kd \right|^2 \right) \notag \\ & +\frac{1}{2} \Gamma_A
 \left( \left|\psi_\ku \right|^4 + \left| \psi_\kd \right|^4 \right)
 +\Gamma_B \left| \psi_\kd \right|^2 \left| \psi_\ku \right|^2 \notag
 \\ & +\frac{1}{2} \Gamma_C \left(\psi^{\ast 2}_\ku \psi^2_\kd
 +c.c. \right),
\label{Energy2}
\end{align}
where $\Gamma_A= \Gamma_{\OO, \ku, \ku}=\Gamma_{\OO, \kd, \kd}=3.36
J$,  $\Gamma_B= \Gamma_{\OO, \ku, \kd}+\Gamma_{\vd, \ku, \kd}=0 $ and
$\Gamma_C= \Gamma_{\vd, \ku, \ku}=1.30 J$. The
amplitude $\Gamma_B$ is zero because the denominator
in Eq.~(\ref{Gamma}),
$\epsl_{\k+\qp}+\epsl_{\kp-\qp}-\epsl_{\k}-\epsl_\kp$, 
depends on only two components of $\qp$ for $\k=\ku $ and $\kp= \kd$.
This leads to vanishing interactions, as is familiar 
from the example of the two dimensional Bose gas.\cite{Fisher01}

The energy of this commensurate state is minimized by setting
$\psi_{\ku}= \pm i \psi_{\kd}$, giving
\be \frac{E}{N}= -\frac{ \left(\epsl_{\ku} -\mu
\right)^2}{\Gamma_A+\Gamma_B-\Gamma_C}\,.   \ee
We find that it is lower than the energy of the incommensurate cone state for fields
smaller than $\hh^c_{T1} = 0.958 \hh^c_{cr}$: see Fig.~\ref{Figure4}.
The proximity of $\hh^c_{T1}$ to $\hh^c_{cr}$ justifies our use of the
low density approximation.
The components of the ordered moment in this commensurate state are
\begin{align}
\langle S^{a}_\r \rangle &=\sqrt{n_c} \left(\cos(\ku \cdot \r +\alpha ) \pm \sin(\kd
\cdot \r +\alpha ) \right), \notag \\ \langle S^{b}_\r\rangle &=\sqrt{n_c}
\left(\sin(\ku \cdot \r+ \alpha ) \mp \cos(\kd \cdot \r+\alpha )
\right), \notag \\ \langle S^{c}_\r\rangle &= \frac{1}{2} - 2 n_c \,,
\label{SC2}
\end{align}
where $n_c=({\mu-\epsl_\ku})/({\Gamma_A+\Gamma_B-\Gamma_C})$ and
$\alpha$ is an arbitrary phase.  This
structure is illustrated in Fig.~\ref{Figure3}.  In this commensurate state
neighboring spins are
antiferromagnetically aligned along the chains and in adjacent layers, but 
perpendicular in adjacent chains. The $\pm$ sign in
Eq.~\ref{SC2} reflects the two distinct ways to arrange this
perpendicular orientation. 
The experimental phase diagram in longitudinal field is summarized
in Fig.~\ref{Figure4}, together with the theoretical results obtained
here for behavior close to the saturation field, and earlier ones for large $S$.


\subsection{Transverse fields}

\begin{figure}[ht]
\begin{center}
\includegraphics[width=8.6cm]{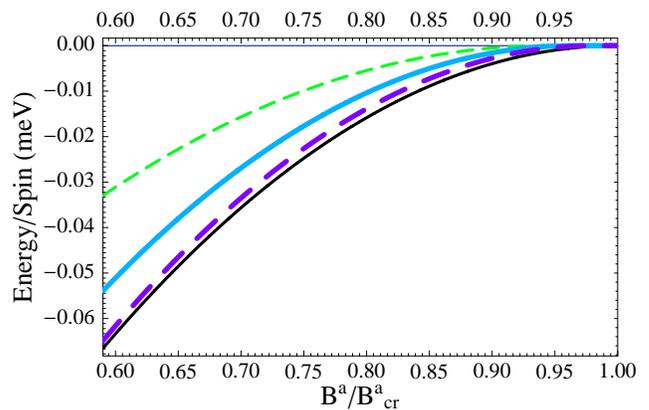}
\caption{(Color Online) Ground state in transverse fields. 
Energies of various states as a function of transverse magnetic field
$B^a$ (normalized with respect to the critical field). 
The thin, thick, short-dashed and long-dashed lines represent,
respectively: the incommensurate cone state; the commensurate
state with condensates at $\ku$ and $\kd$; the incommensurate spin-fan; and the commensurate
state with condensate at $\k=\QC$.}
\label{Figure5}
\end{center}
\end{figure}

In  a transverse field, the DM vector is {\em
parallel} to the field direction and so the DM interaction
is {\em quadratic} in boson creation and annihilation operators. 
Using the standard unit cell containing two layers, and
introducing separate species of bosons for the even and odd layers, the
quadratic Hamiltonian is\cite{Coldea2}
\begin{align}
\ham \!  &= \! \sum_\k \!
\begin{bmatrix}
\phid_{e\k} \! & \!  \phid_{o\k}
\end{bmatrix}
\begin{bmatrix}
 \Omega_\k  \! - \! J^{\prime \prime} \! + \! D_\k \! \! \!  & \! J^{\prime
\prime}_{-\k} \! \notag \\
\! J^{\prime \prime}_\k \! &\!  \! \!  \Omega_\k  \! - \! J^{\prime \prime}
\! - \! D_\k   \\
\end{bmatrix}
\begin{bmatrix}
\phipd_{e\k} \notag \\ \phipd_{o\k}
\end{bmatrix},
\end{align}
where $\Omega_\k= J_\k - J_\OO+ \hh^a $, evaluated with $J^{\prime
\prime}=0$, $D_\k= - D \left(\sin (\k \cdot \d_1)+\sin (\k \cdot \d_2)
\right)$ and $J^{\prime \prime}_\k= J^{\prime \prime} \cos (\k \cdot
\d_3)$.  Diagonalization of this Hamiltonian yields a dispersion relation
with two branches $\epsilon^{\pm}_{\k}$.
Because of DM interactions, the branches lack
inversion symmetry:  $\epsilon^{\pm}_\k \neq \epsilon^{\pm}_{-\k}$.~\cite{Coldea4}
Below the saturation field, order appears at the
incommensurate wavevectors that minimize $\epsilon^{+}_\k$
and  $\epsilon^{-}_\k$.\cite{Coldea2,Veillette01,Coldea4} 
It is likely that any further transition to a commensurate phase
will be disfavored by the reduced symmetry of the dispersion relation
in a transverse field. To test this, we have carried out calculations
of the energies of commensurate states in a transverse field
analogous to the ones described for a longitudinal field.
To simplify these calculations, we have treated a model in
which the DM vector is not staggered between layers, so that the
unit cell contains only a single layer, rather than two.
(See Ref.~\onlinecite{Coldea4} for an alternative approach.)
A similar calculation to the one performed in longitudinal fields reveals  
that the incommensurate state is lowest in energy in transverse
fields (see Fig. \ref{Figure5}) in agreement with experimental data and the $1/S$
expansion. 

\section{Summary}

In summary, we have studied the magnetic ground states of
\cscucl in the vicinity of the saturation field by treating reversed spins
as a dilute Bose gas. In a transverse
field we find only an incommensurate ground state ,
in agreement with experiment and calculations based
on the $1/S$ expansion. By contrast, in a 
longitudinal field there is a transition
between an incommensurate state close to the saturation field
and a commensurate state at lower field. 
We propose that this commensurate state is the one
recently observed in neutron diffraction.\cite{Unknown01}

\section*{Acknowledgments}

We are grateful to R. Coldea for many discussions and for
sharing experimental results prior to publication. 
This work was
supported by EPSRC under Grant GR/R83712/01.




\begin{thebibliography}{99}


\bibitem{Coldea00} R. Coldea, D. A. Tennant, R. A. Cowley, 
D. F. McMorrow, B. Dorner, and Z. Tylczynski, J. Phys.: Condens. Matter
\textbf{8}, 7473 (1996).

\bibitem{Coldea1} R. Coldea, D. A. Tennant, A. M. Tsvelik, and
Z. Tylczynski, Phys. Rev. Lett. \textbf{ 86}, 1335 (2001).

\bibitem{Coldea2} R. Coldea, D. A. Tennant, K. Habicht, P.
Smeibidl, C. Wolters, and Z. Tylczynski, Phys. Rev. Lett.
\textbf{88}, 137203 (2002).

\bibitem{Coldea3} R. Coldea, D. A. Tennant, and Z. Tylczynski, 
Phys. Rev. B \textbf{68}, 134424 (2003).

\bibitem{Coldea4} T. Radu, H. Wilhelm, V. Yushankhai, D.
Kovrizhin, R. Coldea, Z. Tylczynski, T. Luehmann, and F. Steglich,
Phys. Rev. Lett. \textbf{95}, 127202 (2005).

\bibitem{Unknown01} R. Coldea and D.A. Tennant (private communication).

\bibitem{Anderson01} P.W. Anderson, Mater. Res. Bull. \textbf{8}, 153 (1973).

\bibitem{Bocquet01} M. Bocquet, F. H. L. Essler, A. M. Tsvelik, and
A. O. Gogolin, Phys. Rev. B \textbf{ 64}, 094425 (2001).

\bibitem{Chung01} C. H. Chung, J. B. Marston, and R. H.
McKenzie, J. Phys.: Condens. Matter \textbf{13}, 5159 (2001).

\bibitem{Chung02} C. -H. Chung, K. Voelker, and Y. B. Kim, Phys.
Rev. B \textbf{ 68}, 094412 (2003).

\bibitem{Isakov01} S. V. Isakov, T. Senthil, and Y. B. Kim, 
Phys. Rev. B \textbf{ 72}, 174417 (2005).

\bibitem{Veillette02} M. Y. Veillette, A. J. A. James, and F. H. L.
Essler, Phys. Rev. B \textbf{ 72}, 134429 (2005).

\bibitem{Alicea01} J. Alicea, O. I. Motrunich, and M. P. A. Fisher, (unpublished), cond-mat/0508536.

\bibitem{Zheng03} W. Zheng, J. O. Fjaerestad, R. R. P. Singh, 
R. H. Mackenzie, and R. Coldea, (unpublished), cond-mat/0506400.

\bibitem{Veillette01} M. Y. Veillette, J. T. Chalker, and R. Coldea, 
Phys. Rev. B \textbf{ 71}, 214426 (2005).

\bibitem{Kovrizhin} D. L. Kovrizhin, V. Yushankai, and L. Siurakshina,
(unpublished) cond-mat/0509552.

\bibitem{Comment1} Experimentally, there appears to be a weak anisotropy {\em within}
the easy plane, which is not included in the model Hamiltonian, and which we
ignore in this paper. 



\bibitem{Batyev01} E. G. Batyev and L. S. Braginskii,
Zh. Eksp. Teor. Fiz. \textbf{ 87}, 1361 (1984) [Sov. Phys. JETP \textbf{ 60},
781 (1984)]. 

\bibitem{Batyev02} E. G. Batyev, Zh. Eksp. Teor. Fiz. \textbf{ 89}, 308
(1985) [ Sov. Phys. JETP \textbf{ 62}, 173 (1985)].

\bibitem{Gluzman01} S. Gluzman, Z. Phys. B: Condens. Matter \textbf{ 90}, 313, (1993).

\bibitem{Nikuni01} T. Nikuni and H. Shiba, J. Phys. Soc. Jpn. 
\textbf{ 64}, 3471 (1995).




\bibitem{Matsubara01} T. Matsubara and H. Matsuda, Prog. Theor.
Phys. \textbf{ 16}, 569 (1956).

\bibitem{Abrikosov} A. A. Abrikosov, L. P. Gorkov, and I. E.
Dzyaloshinskii in {\em Methods of Quantum Field Theory In Statistical
Physics}, (Dover, New York, 1975).

\bibitem{Beliaev01} 
S. T. Beliaev, Zh. Eksp. Teor. Fiz. \textbf{ 34}, 
417 (1958) [Sov. Phys. JETP \textbf{ 7}, 289 (1958)].
S. T. Beliaev, Zh. Eksp. Teor. Fiz. \textbf{ 34}, 
433 (1958) [Sov. Phys. JETP \textbf{ 7}, 299 (1958)].

\bibitem{Fisher01} D. S. Fisher and P. C. Hohenberg, Phys.
Rev. B \textbf{ 37}, 4936 (1988). 


\bibitem{Gfactor} S. Bailleul, J. Holsa, and P. Porcher, Eur. J.
Solid State Inorg. Chem. \textbf{ 31}, 432 (1994).
 










\end{thebibliography}
\end{document}